\documentclass[journal]{IEEEtai}[2007/03/05 V1.7a by Michael Shell]

\usepackage[colorlinks,urlcolor=blue,linkcolor=blue,citecolor=blue]{hyperref}

\usepackage{color,array}
\usepackage{cite}
\usepackage{amsmath,amssymb,amsfonts}
\usepackage{algorithmic}
\usepackage{graphicx}
\usepackage{graphicx}
\usepackage{amsmath}
\usepackage{amssymb}
\usepackage{booktabs}
\usepackage{multirow}
\usepackage{graphicx}
\usepackage{multirow}
\usepackage{graphicx,multirow}
\usepackage{bm}
\usepackage{hyperref}
\usepackage{adjustbox}
\usepackage{graphicx}
\usepackage{textcomp}
\usepackage{graphicx}
\usepackage{tikz}
\usepackage{graphicx}
\usepackage{tabularx}
\usepackage{longtable}
\usepackage{supertabular}
\usepackage{newfloat}
\usepackage[percent]{overpic}
\usepackage{xspace}
\usepackage{newfloat}
\usepackage{tikz}
\usetikzlibrary{arrows}
\usetikzlibrary{shapes}
\usepackage{nicefrac}
\usepackage{hyperref}
\usepackage{orcidlink}

\begin{document}

\title{Improving  3D Medical Image Segmentation at Boundary Regions using Local Self-attention and Global Volume Mixing}
\author{Daniya~Najiha~Abdul~Kareem \orcidlink{0000-0001-9788-7688}, Mustansar~Fiaz \orcidlink{0000-0003-2289-2284}, Noa~Novershtern, Jacob~Hanna \orcidlink{0000-0003-2042-9974}, and Hisham~Cholakkal
\thanks{Daniya Najiha Abdul Kareem, Mustansar Fiaz and Hisham Cholakkal are with the Mohamed bin Zayed University of Artificial Intelligence, UAE.}
\thanks{Noa Novershtern and Jacob Hanna   are with the Weizmann Institute of Science, Israel.}}

\markboth{Journal of IEEE Transactions on Artificial Intelligence, Vol. 00, No. 0, Month 2020}
{Daniya~Najiha~Abdul~Kareem \MakeLowercase{\textit{et al.}}: vMixer}

\maketitle

\begin{abstract}
Volumetric medical image segmentation is a fundamental problem in medical image analysis 
where the objective is to accurately classify a given 3D volumetric medical image with voxel-level precision. In this work, we propose a novel hierarchical encoder-decoder-based framework that strives to explicitly capture the local and global dependencies for volumetric 3D medical image segmentation. The proposed framework exploits local volume-based self-attention to encode the local dependencies at high resolution and introduces a novel volumetric MLP-mixer to capture the global dependencies at low-resolution feature representations, respectively. The proposed volumetric MLP-mixer learns better associations among volumetric feature representations. These explicit local and global feature representations contribute to better learning of the shape-boundary characteristics of the organs. 
Extensive experiments on three different datasets reveal that the proposed method achieves favorable performance compared to state-of-the-art approaches. On the challenging Synapse Multi-organ dataset, the proposed method achieves an absolute 3.82\% gain over the state-of-the-art approaches in terms of HD95 evaluation metrics {while a similar improvement pattern is exhibited in 
MSD Liver and Pancreas tumor datasets}.  We  also provide a detailed comparison between recent  architectural design choices in the 2D computer vision literature by adapting them for the problem of 3D medical image segmentation.  Finally, our experiments  on the ZebraFish 3D cell membrane dataset having limited training data demonstrate the superior transfer learning capabilities of  the proposed vMixer model on  the challenging 3D cell instance segmentation task, where accurate boundary prediction plays a vital role in distinguishing individual cell instances. 
Our source code is publicly available at \url{
https://github.com/Daniyanaj/vMixer}.

\end{abstract}

\begin{IEEEImpStatement}
Automatic medical image segmentation is a crucial step in the healthcare systems to perform accurate diagnoses, and pixel-wise estimation of cancerous tissues and organs from diverse medical images such as CT, MRI, and more. In 3D volumetric segmentation, transformers and hybrid approaches are exposed to self-attention, struggle to learn the inherent complex boundaries of the tissue and exhibit quadratic complexity with the number of tokens. The proposed vMixer framework exploits explicit local and global volumetric features to better learn the {shape-boundary details of the organs}. We also provide an extensive study on the selection of architectural design that is adapted for 3D medical segmentation from 2D vision literature for better boundary localization. Finally, we exploit the transfer learning capabilities of the proposed vMixer where training data is limited. We hope that our contributions can facilitate the research community to make use of artificial intelligence in designing a 3D image segmentation framework.

\end{IEEEImpStatement}


\section{Introduction} \label{sec:introduction}


In clinical diagnosis, volumetric  segmentation is a fundamental task that has shown promising potential in a wide range of applications including  organ localization \cite{isensee2018nnu,xu2019efficient} and tumor identification \cite{myronenko20193d, chen2019s3d}.  UNet \cite{ronneberger2015u} is a breakthrough volumetric medical image segmentation approach that utilizes a CNN-based encoder and decoder architecture, where the encoder generates hierarchical low-dimensional features and decoder maps learned features into a voxel-wise segmentation.   However, UNet \cite{ronneberger2015u}  and its variants including  V-Net \cite{milletari2016v} and  ESPNet \cite{nuechterlein20193d}, often struggle to capture the long-range feature dependencies due to the limited receptive field of the convolution operation.   This is particularly problematic in the case of multi-organ segmentation having large variations in the shapes and scales of organs. 

Recently, transformers-based  \cite{swin-unet} approaches have been explored to capture global feature dependencies, which are further improved by hybrid approaches \cite{miss,nnformer,fiaz,kareem} that leverage the benefits of self-attention along with CNN components. Although self-attention learns the global pair-wise dependencies from volumetric 3D medical data, it struggles to capture the underlying complex boundaries of the tissues. Moreover, standard self-attention operates on pairwise patches which have quadratic complexity with respect to the number of tokens.
 {In this work, our objective is to provide a hybrid architecture that helps to learn local as well as global representation to capture better shape-boundary information of the complex organs (shape of the organs). }
Motivated by the success of Multi-Layer Perceptron-mixers (MLP-mixers) \cite{tolstikhin2021mlp} for image classification in natural images,  we propose a novel volumetric medical image segmentation approach that strives to reduce the segmentation errors 
by introducing  MLP-mixer and window-based self-attention blocks to explicitly capture global and local dependencies of volumetric feature representations, respectively.  

Generally, the performance of volumetric segmentation approaches is evaluated using dice similarity coefficient (DSC), Hausdorff distance (HD95), and Normalised surface distance (NSD) metrics. DSC measures the overlap index  between the ground-truth and predicted segmentation masks and NSD measures the overlap between the segmented boundaries.  On the other-hand,  Hausdorff distance (HD95) is measured as a distance between the boundaries of predicted and ground-truth segmentations \cite{karimi2019reducing},\cite{taha2015metrics},   hence, higher values for the DSC score and NSD score, and a lower HD95 score indicates better model performance.  Among these metrics,  HD95 is the most informative and useful criterion as it indicates the largest segmentation error, especially where organs have varying sizes and shapes. Moreover, in applications where segmentation is an initial step in a complex multi-step process, the largest segmentation error evaluated in the HD95 score is a good indicator of the usefulness of segmentation for the given application \cite{karimi2019reducing}.
Despite this, state-of-the-art volumetric segmentation approaches are sub-optimal in their HD95 evaluation metric.  

Transfer learning has shown great progress in various areas including computer vision \cite{guo2019spottune}, natural language processing \cite{sung2022vl}, medical image segmentation \cite{chen2019med3d}, and remote sensing \cite{cui2020semantic}. Transfer learning provides a powerful technique to benefit from the existing knowledge and to improve the efficiency and effectiveness of the target application and domain. Therefore, in this paper, we also explore  different convolutional-based, transformer-based, mixer-based, and hybrid-based architectural design choices to get leverage from transfer learning where the training data is limited. Our experimental study reveals that the proposed hybrid architectural design exhibits better transfer learning compared to other architectural designs.

\subsection{Contributions}
  {\textbf{(i)} In this work, we propose a hybrid hierarchical encoder-decoder framework, termed vMixer,  that strives to capture both local and global information for accurate boundary prediction during volumetric medical image segmentation. }
  {Based on our comprehensive studies, we propose a novel  MLP mixer-based framework for medical image segmentation that utilizes volume-based self-attention (Swin attention) to capture local dependencies at the high-resolution stage and introduces a novel Global Volume Mixer (GVM) block to encode the global dependencies at lower-resolution stages.  This explicit utilization of global and local representation leads to better learning of organ boundary regions. }

 {\textbf{(ii)} We perform a comprehensive comparison between different architectural design choices available in the literature by adapting them for 3D medical image segmentation. To the best of our knowledge, we are the first to have such a comprehensive study, and we hope that our study will support future research in this direction.}

 {\textbf{(iii)} Our comprehensive experiments on three different datasets (Synapse Multi-organ, MSD-Pancreas Tumour, and  MSD-Liver Tumour) across HD95, Dice, and NSD evaluation metrics show the merits of the proposed approach. 
In addition, our experiments on  ZebraFish 3D cell membrane dataset (with limited training data) show that the proposed vMixer exhibits superior transfer learning abilities for challenging 3D cell instance segmentation task where accurate boundary prediction is crucial for delineating different cell instances. }

\begin{figure*}[h]
    \centering  
\includegraphics[width=0.9\linewidth]{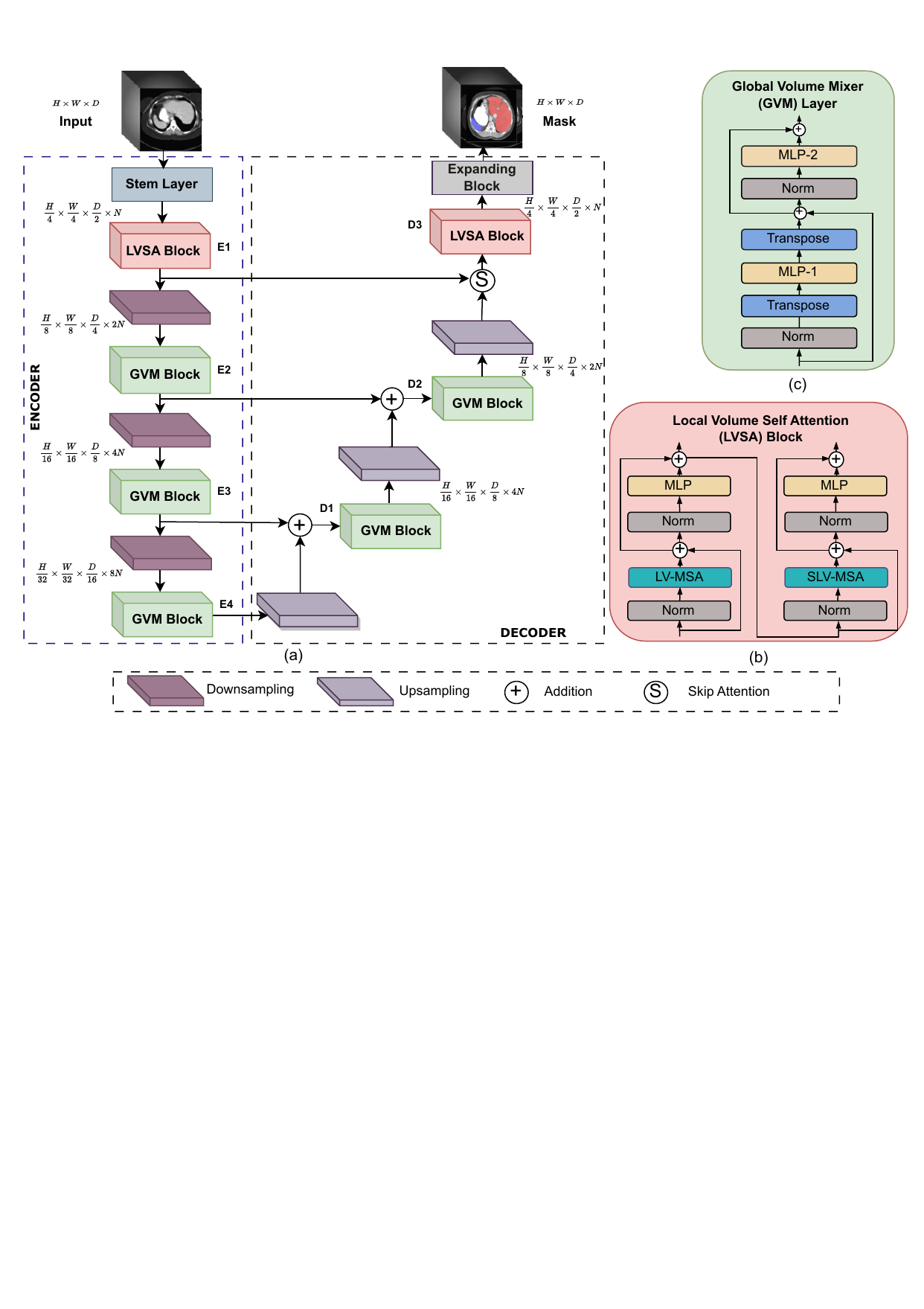}
    \vspace{-10cm}
    \caption{ {(a) Overview of the proposed vMixer framework with hierarchical encoder-decoder architecture. The focus of our design is to explicitly capture the local and global feature dependencies for accurate segmentation.  Our framework takes  3D images as input and employs local volume self-attention (LVSA) block 
 to explicitly learn the local dependencies at high resolution ($E1$, $D3$). The $E1$  features are downsampled and passed to the proposed global mixer block to explicitly learn the global dependencies.   In the decoder, the features are first upsampled and then fused with the encoder features through a skip connection. 
 We employ global volume mixer blocks at the first two decoder stages ($D1$ and $D2$) and a LVSA block at the last stage of the decoder ($D3$).  The final decoder features are fed to an expanding layer for producing the final segmentation mask.
 (b) Presents the  LVSA block which comprises of local volume-based multi-head self-attention (LV-MSA) layer followed by a shifted local volume-based multi-head self-attention  (SLV-MSA) layer.  
 (c) Shows the structure of the volumetric MLP-mixer layer used in the GVM  block. Each GVM block comprises two MLP-mixer layers. The volumetric MLP-mixer layer performs token mixing and channel mixing operations on the input volumetric tokens. 
 }}
\label{fig:overall_proposed_framework}
 \end{figure*}
 
\section{Related Work}
3D volumetric segmentation helps in identifying the regions of interest within a 3D volume and aids in diagnostic imaging, treatment evaluation, and  surgical planning by accurate delineation of anatomical structures or abnormalities. Numerous studies have been evidenced by the literature which encompasses a wide range of approaches among which CNN-based models, transformer-based models, and hybrid models remain the most prominent ones. U-Net proposed by Long et al.\cite{2D} has gained popularity over the last decade and is widely utilized in image segmentation methods. These models, built upon the U-Net framework, continue to be at the forefront of cutting-edge design among the image segmentation algorithms incorporating both CNNs and transformers.

\subsection{Convolutional Neural Networks Based Models}
Convolutional Neural Networks (CNNs) have gained significant popularity in 3D medical segmentation due to their ability to process volumetric data to effectively capture and learn hierarchical features. Multiple convolutional layers in CNNs help to exploit local spatial dependencies within the 3D data. 
Milletari et al.\cite{3dunet} introduced a 3D U-Net architecture that incorporates residual blocks instead of cascaded CNNs, and pooling layers were replaced by strided convolutions. Additionally, a loss function based on the dice score was employed to address the issue of class imbalance among voxel groups. By leveraging these adaptations, the 3D U-Net architecture improved the accuracy and effectiveness of 3D medical segmentation.
Rehman et al. \cite{rehman2023maxvit} extended the MaxVit \cite{tu2022maxvit} into the UNet architecture and used it for 2D cell segmentation tasks similar to \cite{aralikatti2023dual, saif2021capscovnet, mahmud2021covsegnet}.
Peng et al. \cite{peng2020multi} introduced a variant of the 3D U-Net architecture which involved utilizing multiple U-Net modules for the extraction of long-range spatial details at different scales. Xception blocks were employed instead of conventional convolutions within the U-Net blocks to enhance the feature extraction.  { Moreover, Peng et al.} utilized 3D convolutions with depthwise separable convolutions aiming to reduce the computational complexity. The nn-UNet framework proposed in \cite{isensee2018nnu} is another noteworthy contribution to the field of 3D medical image segmentation. This framework is essentially built upon three simple U-Net models and has introduced many optimal strategies for pre-processing, training, inference computations, and post-processing  that greatly contribute to effective network implementation. Since its introduction, the nn-UNet framework has been widely used in the latest image segmentation methods.
 

The resolution of 3D images is a major cause of concern which often demand substantial computational resources while training a 3D model. Recently, patch-based segmentation methods have been developed to address the issue. Kamnitsas et al. \cite{Kamnitsas} proposed a  dual-network architecture for brain-lesion segmentation that  captures information from two distinct receptive fields and simultaneously learns features at different scales. In this model, dense fully connected layers are utilized to classify voxels into different groups from patchified input images. With this approach, the accuracy and efficiency of brain lesion segmentation in 3D medical images have shown remarkable improvement compared to other SOTA methods. 

\subsection{Transformer Based Models}
Transformers have demonstrated exceptional performance in the analysis of volumetric medical images by effectively modeling sequential data and encoding long-range dependencies. Transformers excel in capturing global interactions and contextual information across the entire 3D volume, unlike traditional CNN-based approaches that mainly focus on local spatial relationships \cite{fang2022annotation, karimi2022improving}. 
Karimi et al. \cite{karimi} introduced a transformer-based model manipulating the self-attention mechanism between consecutive linear embeddings of image patches, enabling the model to effectively capture spatial relationships and dependencies. Additionally, an effective pre-training method was also adopted for the model, which proved valuable in scenarios where only limited annotated data is available. Experiments conducted on three different medical 3D datasets, including the hippocampus, pancreas, and brain cortical plate demonstrated the  model's efficiency.

Swin Transformers introduced by Liu  et. al \cite{swin} reduces the complexity linearly with the number of tokens compared to the quadratic computational complexity in ViTs \cite{vaswani2017attention}  by utilizing window-based multi-scale attention. In addition, it allows a variable token size which enables them to handle objects of variable scales commonly found in medical images. Swin Transformers capture both local and global information effectively and also incorporate hierarchical feature extraction through patch merging layers and repeated Swin Transformer blocks at different image scales.
Cao et al. \cite{swin-unet} introduced Swin-UNet, an architecture that integrates Swin transformer blocks into the traditional U-Net framework in an encoder-decoder design. The Swin-UNet architectural design achieved promising segmentation accuracies for Synapse and ACDC datasets.

\subsection{Hybrid Architectures}
In recent studies, the dominant approach utilizes  U-shaped architectures incorporating transformers and CNNs through various strategies incorporating multi-scale feature extraction techniques and self-attention layers in the network \cite{khan2023survey}. The excellence of CNNs in capturing local spatial features and extracting hierarchical representations makes them well-suited for encoding fine-grained details and local patterns within the 3D volume. Transformers, on the other hand, have the ability to model global dependencies and encompass long-range contextual information. As identification of the complex relationships between different regions and structures within the volume is important in 3D medical image segmentation, hybrid architectures play a significant role.

The TransUNet was proposed by Chen et al.\cite{transunet} which utilizes a transformer for global feature encoding and CNN for high-resolution feature extraction. Transformer features are added with skip connections from the encoder network for precise localization. This network achieved better performance in organ segmentation compared to ViTs. In \cite{unetr},
UNETR was designed by Hatamizadeh et.al. based on ViTs \cite{vaswani2017attention} that includes a transformer encoder and a CNN decoder which is connected using skip connections. This network achieves good performance on MSD and BTCV segmentation datasets. In another approach, 
Hatamizadeh et al. \cite{swinunetr} introduced Swin UNETR, an architecture based on Swin UNet, incorporating Swin Transformers in the encoder and utilizing a CNN decoder. This model stands out as one of the top-performing approaches in the BraTs2021 MRI dataset and Synapse multi-organ CT dataset. With an efficient window partitioning scheme and attention mechanism, Swin UNETR showcases its capabilities in achieving accurate segmentation results.

Many of the aforementioned studies focused on encoding global context by incorporating transformers along with CNN as supplementary modules but they did not fully optimize the integration of convolution and self-attention operations. To overcome this limitation, Zhou et al. introduced nnFormer \cite{nnformer}  within the nn-UNet \cite{isensee2018nnu} framework, presenting an interleaved architecture combining both convolution and self-attention mechanisms. nnFormer generates hierarchical features and expands the receptive fields by effectively leveraging both local volume-based and global volume-based self-attention mechanisms. 
Compared to existing methods, we propose a hybrid architectural design that benefits in the extraction of local details using self-attention at high-resolution and global dependencies using MLP-Mixer at low-resolution features for volumetric 3D medical image segmentation.  This combination handles the complexity associated with 3D image segmentation tasks and helps in achieving an optimal design choice.  We also leverage the transfer learning capabilities of our design over a 3D cell membrane dataset.


\section{The proposed Method}
\noindent\textbf{Motivation:} 
As discussed earlier, state-of-the-art transformer-based and hybrid approaches generally employ self-attention operations to obtain high-quality segmentation results. However, these approaches often struggle to predict accurate organ boundaries. 
Here,  we argue that it is desired to  learn the  boundary regions of the organs occurring in the local and larger spatial context. To this end,  it is desired to analyze different architectural choices in the 2D  computer vision literature by adapting them for 3D medical image segmentation.  
In this work, we  explicitly learn local dependencies at high-resolution features while capturing global dependencies at lower-resolution features. This approach benefits from learning better associations among the volumetric feature representation, which leads to better prediction of organ boundary regions.  

\subsection{Overall Architecture}
Fig. \ref{fig:overall_proposed_framework}-(a) shows our overall hierarchical encoder-decoder framework.
The proposed framework has four encoder and three decoder stages.
 {As discussed earlier, accurate organ boundary segmentation is a complex task that requires local contextual information for the precise delineation of boundary pixels from the background, while it simultaneously requires global contextual information to prevent erroneous predictions. Therefore, to learn shape-boundary information about the varying shapes of organs, we propose to explicitly capture local and global dependencies.} To be specific, we capture local feature dependencies at the first encoder and last decoder stages having the highest feature resolution, and global feature dependencies are captured at the remaining encoder and decoder stages having relatively lower feature resolutions. 
 {
The first stage of the encoder consists of the stem layer followed by the local volumetric self-attention (LVSA) block which comprises local volume-based multi-head self-attention and shifted local volume-based multi-head self-attention layers. While the latter three encoder stages are composed of a downsampling layer followed by a global volumetric mixer (GVM) block (which has two global volumetric MLP-mixer layers). GVM block at low-resolution features exhibits a holistic approach which helps in better extraction of global detail that can possibly learn complex organ shapes in the volumetric context by performing token mixing. } Similar to the encoder, the decoder also follows a multi-stage hierarchical architecture.  Each decoder stage utilizes upsampling layer to increase the feature resolution followed by a GVM block. Finally, in the last stage of the decoder, we follow a structure similar to the first encoder stage i.e.,  we employ a LVSA block followed by a feature-expanding layer that predicts the final masks.  




\subsection{Local Volume Self Attention}
Our encoder takes 3D input images as input to the stem layers. These stem features and last-stage decoder features have high-resolution features. Hence,  applying self-attention on uniformly sampled dense patches from these high-resolution feature maps leads to a quadratic complexity with respect to the number of tokens. To learn the explicit local dependencies, we adopt a local volume-based self-attention block instead of 2D  local windows as in Swin Transformer \cite{liu2021swin}, which has reduced computational complexity compared to standard self-attention. Similar to the Swin Transformer block, we endorse local feature encoding with the help of a local volume-based multi-head self-attention layer followed by a locally shifted volume-based multi-head self-attention layer as shown in Fig. \ref{fig:overall_proposed_framework}-(b).

\subsection{Global Volume Mixing}
As mentioned earlier, to learn the complex shapes of the organs in the volumetric context, we propose a global volume mixer block to explicitly capture the global dependencies from low-resolution stages. Standard self-attention \cite{dosovitskiy2020image} operates over dense patches that have quadratic complexity with respect to the total number of tokens. 
 {MLP-Mixer \cite{tolstikhin2021mlp} can possibly learn complex relationships across the
entire input, which makes them effective at capturing global information by performing token mixing followed by a pointwise feature refinement. 
In contrast to other context aggregators \cite{kaiser2017depthwise,vaswani2017attention}, MLP-mixer is more dense, static, and does not require parameter sharing \cite{gao2021container}. 
The core operation of MLP-mixer is the dense transposed affinity matrix on a single feature group. Therefore, we introduce a global volume mixer (GVM) block that has two volumetric MLP-mixer layers to capture the global dependencies for the underlying feature representation to better learn the complex boundaries of the tissues}. The volumetric MLP-mixer layer is composed of layer norm, transposed token-mixing, and a MLP with fully-connected layers with a GELU nonlinearity, as shown in Figure \ref{fig:overall_proposed_framework}-(c).

 {Suppose $\mathcal{F} \in  \mathcal{R}^{H\times W\times D\times N}$, 
 which is reshaped to $\mathcal{F} \in \mathcal{R}^{M \times N}$  and taken as input to volumetric global MLP-mixer layer, where $M=(H \times W \times D)$ represents the size of the 3D input (volume) and $N$  denotes the number of channels.}
The volumetric global MLP-mixer operations can be summarized as:

\begin{equation}
\begin{aligned}
    \mathcal{\hat{F}}=( W_{mlp-1}(\text{Norm} (\mathcal{F} )^T))^T+\mathcal{F},\\
    \mathcal{\bar{F}}= W_{mlp-2}(\text{Norm} (\mathcal{\hat{F}})+\mathcal{\hat{F}},\\
\end{aligned}
\end{equation}
where $W_{mlp-1}$ and $W_{mlp-2}$ denote the learnable multi-perceptron layer weights and  $\hat{F}$ and  $\bar{F}$ represent the intermediate and final volumetric global mixing features, respectively.

\begin{table}[t]
\centering
\caption{ {Comparison with other state-of-the-art methods over Synapse Multi-organ dataset. The best results are in bold. }}
\scalebox{1.2}{
        \begin{tabular}{|l|c|c|c|}
        \hline
       Method    &DSC  & HD95 & NSD \\ \hline \hline
        UNet \cite{ronneberger2015u} & 76.85 & - &-\\ \hline
        ViT \cite{dosovitskiy2020image}+CUP \cite{transunet}                  & 67.86 & 36.11& - \\ \hline
        R50-ViT \cite{dosovitskiy2020image} + CUP \cite{transunet}            & 71.29 & 32.87 &-\\ \hline
        TransUNet \cite{transunet}      & 77.48 & 31.69&- \\ \hline
        SwinUNet \cite{swin-unet}       & 79.13 & 21.55 &-\\ \hline
        MissFormer \cite{miss}   & 81.96 & 18.20 &-\\ \hline
        UNETR \cite{unetr}                    & 79.56 & 22.97 & 85.34\\ \hline
        Swin UNETR \cite{swinunetr}                    & 80.24 & 17.65 & 85.75\\\hline
        nnFormer \cite{nnformer}           & \textbf{86.57}   &10.63 &92.04    \\ \hline
        \textbf{vMixer (Ours)}        & {86.53}     & \textbf{6.78}  &\textbf{92.96}   \\ \hline
        \end{tabular}}
        
\label{tbl_all_synapse}
\end{table}

\begin{table*}
\centering
\caption{Organ-wise segmentation comparison between  UNETR, nnFormer, and our vMixer over Synapse Multi-organ dataset. The best results are in bold.}
\scalebox{0.8}{
\begin{tabular}{|c|cc|cc|cc|cc|cc|cc|cc|cc|cc|}
\hline
\multicolumn{1}{|l|}{\multirow{2}{*}{Methods}} & \multicolumn{2}{c|}{\textbf{Average}}       & \multicolumn{2}{c|}{\textbf{Aorta}}         & \multicolumn{2}{c|}{\textbf{Gall Bladder}}  & \multicolumn{2}{c|}{\textbf{Kidney(L)}}     & \multicolumn{2}{c|}{\textbf{Kidney(R)}}     & \multicolumn{2}{c|}{\textbf{Liver}}        & \multicolumn{2}{c|}{\textbf{Pancreas}}     & \multicolumn{2}{c|}{\textbf{Spleen}}        & \multicolumn{2}{c|}{\textbf{Stomach}}      \\ \cline{2-19} 
                         & \multicolumn{1}{c|}{DSC}    & HD95    & \multicolumn{1}{c|}{DSC}    & HD95    & \multicolumn{1}{c|}{DSC}    & HD95    & \multicolumn{1}{c|}{DSC}    & HD95  & \multicolumn{1}{c|}{DSC}    & HD95    & \multicolumn{1}{c|}{DSC}    & HD95   & \multicolumn{1}{c|}{DSC}    & HD95   & \multicolumn{1}{c|}{DSC}    & HD95    & \multicolumn{1}{c|}{DSC}    & HD95   \\ \hline
UNETR \cite{unetr}   & \multicolumn{1}{c|}{79.5} & 22.97 & \multicolumn{1}{c|}{89.9} & 5.48 & \multicolumn{1}{c|}{60.55} & 28.69 & \multicolumn{1}{c|}{85.66} & 17.76 & \multicolumn{1}{c|}{84.80} & 22.44 & \multicolumn{1}{c|}{94.45} & 30.40    & \multicolumn{1}{c|}{59.24} & 15.82 & \multicolumn{1}{c|}{87.8} & 47.12 & \multicolumn{1}{c|}{73.99} & 16.05 \\ \hline

                         
nnFormer \cite{nnformer}    & \multicolumn{1}{c|}{\textbf{86.57}} & 10.63 & \multicolumn{1}{c|}{\textbf{92.04}} & 11.38 & \multicolumn{1}{c|}{70.17} & 11.55 & \multicolumn{1}{c|}{86.57} & 18.09 & \multicolumn{1}{c|}{86.25} & 12.76 & \multicolumn{1}{c|}{\textbf{96.84}} & \textbf{2.00}    & \multicolumn{1}{c|}{\textbf{83.35}} & \textbf{3.72} & \multicolumn{1}{c|}{\textbf{90.51}} & 16.92 & \multicolumn{1}{c|}{86.83} & 8.58 \\ \hline
\textbf{Ours} & \multicolumn{1}{c|}{86.53} & \textbf{6.78}  & \multicolumn{1}{c|}{90.63} & \textbf{6.13}  & \multicolumn{1}{c|}{\textbf{70.33}} & \textbf{9.04}  & \multicolumn{1}{c|}{\textbf{88.74}} & \textbf{5.60}  & \multicolumn{1}{c|}{\textbf{87.38}} & \textbf{7.25}  & \multicolumn{1}{c|}{96.74} & 2.43 & \multicolumn{1}{c|}{80.34} & 4.53 & \multicolumn{1}{c|}{89.71} & \textbf{12.82} & \multicolumn{1}{c|}{\textbf{87.10}} & \textbf{8.45} \\ \hline
\end{tabular}}
\label{tb_nnunet}
\end{table*}

\begin{figure*}[t]
    \centering
\includegraphics[width=1.0\linewidth]{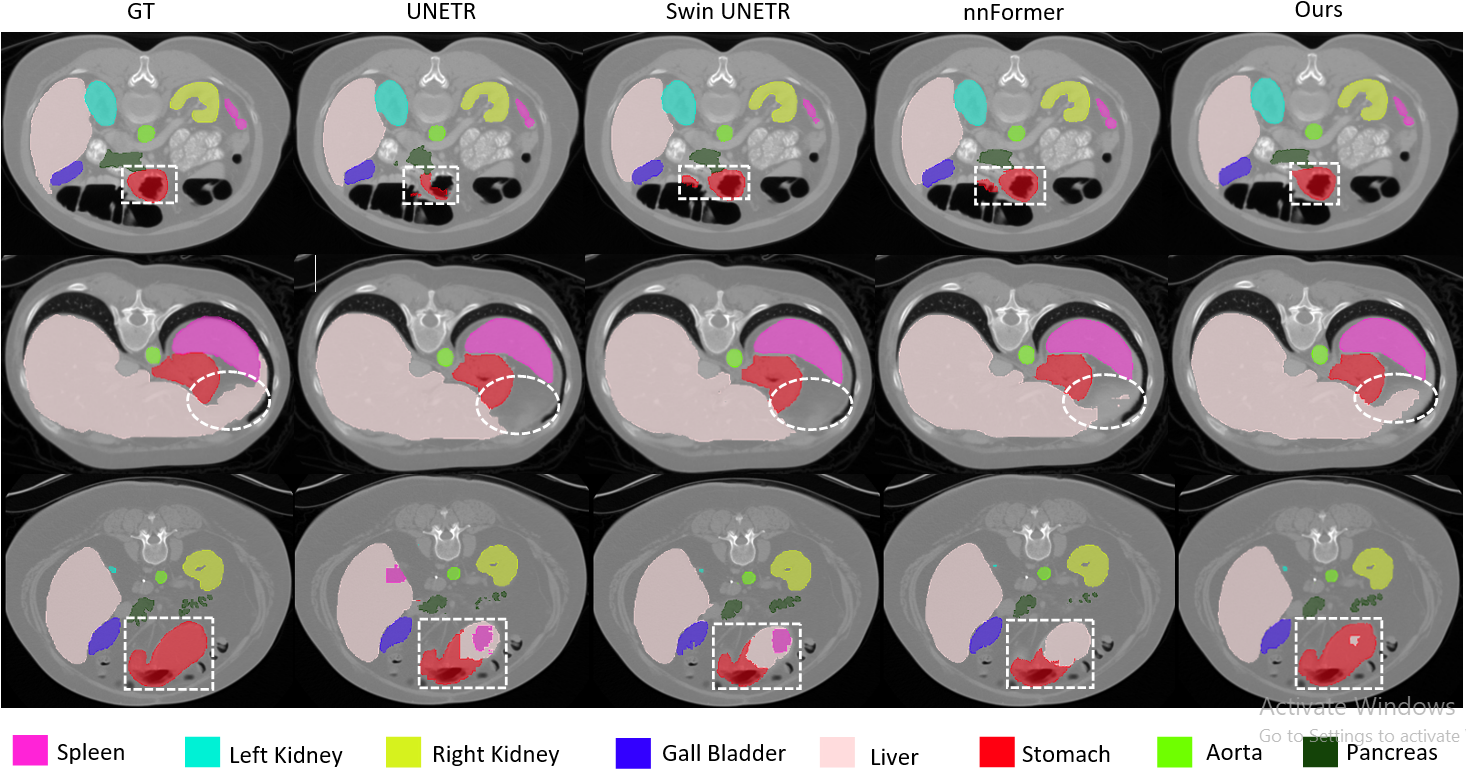}
\caption{Qualitative comparison on the Synapse multi-organ  dataset. Our method provides improved segmentation by accurately detecting the boundaries of the organ. 
}
\label{fig:synapse_all}
\end{figure*}

\begin{figure}[t!]
   \centering
\includegraphics[width=0.5\textwidth]{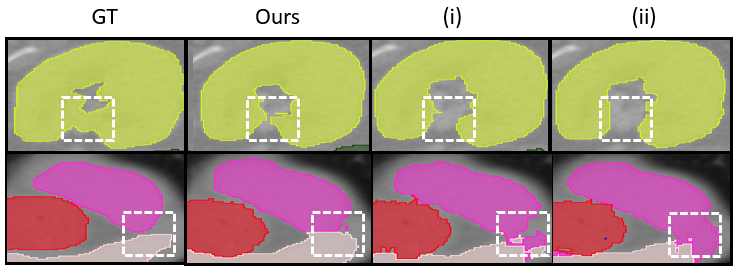}
\caption{ {\textbf{Qualitative comparison} of ablation experiments on Synapse multi-organ dataset. In a closer inspection, it can be seen that our method (LVSA (stage 1) and GVM (stage 2, stage 3, stage 4)) performs better than (i) GVM (Stages 1-4) and (ii) LVSA (Stages 1-4). Row 1 corresponds to the cross-section of the left kidney and row 2 shows a cross-section containing portions of the liver (light pink), spleen (magenta), and stomach (red). It can be clearly observed that our method has better shape preservation capabilities compared to other settings. }}
\label{vis}
\end{figure}

\subsection{Additional Layers}
\noindent\textbf{Stem Layer:} The stem layer is responsible to generate high dimensional tensor  $\mathcal{F}_s \in \mathcal{R}^{\frac{H}{4} \times \frac{W}{4} \times \frac{D}{2}}$ for the  input image $\mathcal{X}\in \mathcal{R}^{H \times W \times D}$, where $(H,W,D)$ is the shape of 3D input (volume). We apply  four successive convolutional layers with kernel size 3 for  tokenization. This reduces the computational complexity compared to the usage of large convolutional kernels as in ViT   \cite{dosovitskiy2020image} and also helps in encoding pixel-level spatial details. Each convolution is followed by GELU activation and layer normalization operations.

\noindent\textbf{Down Sampling:} We perform the downsampling  operation after each stage except the last stage utilizing convolution with a kernel size of 3 with a stride of 2.  This helps in modeling objects at different scales as hierarchical details are obtained by convolution downsampling.

\noindent\textbf{Up-Sampling and Patch Expanding:} In contrast to downsampling in the encoder, up-sampling is performed at each stage of the decoder with the help of convolutional upsampling. A 3D transposed convolution layer is used with a kernel size of 2 and a stride 2 that helps in upsampling the low-resolution feature maps to a higher resolution. In the decoder, these up-sampled feature maps are then added from the encoder to encode the fine-grained details along with the semantic details. At the last stage of the decoder, patch expanding is performed to obtain the final prediction masks using the deconvolutional operation. 

\section {Experiments}

\subsection{Dataset and Evaluation Metrics}
 {We select four 3D medical segmentation datasets for evaluation, with diverse task objectives and versatile granularities. Synapse multi-organ dataset \cite{landman2015miccai} aids for the segmentation of multiple organs while MSD Liver \cite{Antonelli_2022} and MSD Pancreas \cite{Antonelli_2022} datasets provide data for segmenting respective organs and tumors growths over it. On the other hand, ZebraFish Cell Membrane Dataset \cite{cellseg} is a cell instance segmentation dataset that has images with numerous cell targets present in it.}

 {
\noindent\textbf{Synapse Multi-organ dataset:} is a multi-organ dataset  \cite{landman2015miccai} having 30 abdominal CT scans has eight organs including the liver, right kidney, left kidney, pancreas, gall bladder, stomach, spleen, and aorta, and we use a train-test split of 18-12 scans with a resolution of 512×512×160. The abdomen CT scans were acquired from a chemotherapy trial for colorectal cancer and ventral hernia study under the supervision of the Institutional Review Board (IRB).

\noindent\textbf{MSD Liver Tumour Dataset:} The Liver Tumour dataset was introduced as a part of the Medical Segmentation Decathlon Challenge \cite{Antonelli_2022}. It contains 131 3D contrast-enhanced CT scan images from patients with primary
cancers and metastatic liver disease, as a consequence of colorectal, breast, and lung primary cancers with a resolution of 512×512×482. Regions of interest include the liver and tumor regions inside the liver region. We selected this dataset due to its challenging aspect where a major label unbalance is present between the ROIs, i.e. between a larger liver region and a smaller tumor region.
The data was collected from IRCAD Hopitaux Universitaires, Strasbourg which contained random samples from the 2017 Liver Tumor Segmentation (LiTS) challenge \cite{Antonelli_2022}.

\noindent\textbf{MSD Pancreas Tumour Dataset:} This dataset consists of 282 3D CT scan (with a resolution of 512×512×96) volumes of patients undergoing
resection of pancreatic masses which is also a subtask of \cite{Antonelli_2022}. The corresponding target ROIs were the pancreas organ and pancreatic mass (cyst or tumor). This dataset was also selected due to the label unbalance between small (tumor), medium (pancreas), and large (background) structures. The data was acquired at the Memorial Sloan Kettering Cancer Center, in New York, US. \cite{Antonelli_2022}\\
\noindent\textbf{ZebraFish Cell Membrane Dataset:}
For transfer learning, we select 3D zebrafish cell dataset provided by the Department of Systems Biology at Harvard Medical School (HMS) that contains data for cell instance segmentation \cite{cellseg}. It comprises 36 images with a resolution of 181×331×160 which has 32 images for training and 4 images for the test. We utilized  Dice (DSC) similarity,  and Jaccard index (JI) as the evaluation metrics for the zebrafish cell dataset. We evaluate the performance using overall accuracy and cell count accuracy. Overall accuracy indicates the mean value of JI and DSC for all the cells, whereas cell count accuracy is represented as the fractions of cells whose JI or DSC is greater than 50\% or 70\%. 
}

\subsection{Training Setup}
The network is implemented using  PyTorch 1.8.0 and trained using an NVIDIA GeForce RTX 3090 GPU. Following  nnFormer \cite{nnformer}, we adopt the pre-processing and augmentation strategies  and set the batch size 2 and  initial learning rate to 0.01. We utilize a poly decay strategy to adjust the learning rate (lr)  as: 
\begin{equation}\label{lrs}
\begin{aligned}  
\text{lr}=\text{initial\_lr} \times (1-\frac{\text{epoch\_number}}{\text{final\_epoch\_number}})^{0.9}.  
\end{aligned}
\end{equation}
During the training, we use outputs from intermediate and final prediction maps and compute the combined soft dice and cross-entropy losses as in \cite{nnformer}. We set the momentum and weight decay as 0.99 and 3e-5 with SGD  optimizer and trained for 1000 epochs with 250 iterations per epoch.   
 The final loss is calculated as,
 \begin{align}
     \begin{split}
         \mathcal{L}_{all} = \alpha_1 \mathcal{L}_{\{H,\ W,\ D\}} + \alpha_2 \mathcal{L}_{\{\frac{H}{4},\ \frac{W}{4},\ \frac{D}{2}\}} + \alpha_3 \mathcal{L}_{\{\frac{H}{8},\ \frac{W}{8},\ \frac{D}{4}\}}.
     \end{split}
 \end{align}
 Here, $\alpha_{\{1,\ 2,\ 3\}}$ refers to weights for losses, and their values  $\alpha_{\{1,\ 2,\ 3\}}$  are reduced by half with respect to reduction in resolution i.e. $\alpha_3 = \frac{\alpha_1}{4}$ and $\alpha_2 = \frac{\alpha_1}{2}$. All weights are finally normalized to one.

We follow nnUnet \cite{isensee2018nnu} and nnFormer \cite{nnformer} for the pre-processing and augmentation strategies. After pre-processing, we obtained crops with a resolution of 128x128x64 for Synapse Multi-organ, 128x128x128 for MSD Liver Tumour, and  40x224x224 for MSD Pancreas Tumour dataset.

\begin{figure*}[h!]
    \centering
\includegraphics[width=\textwidth]{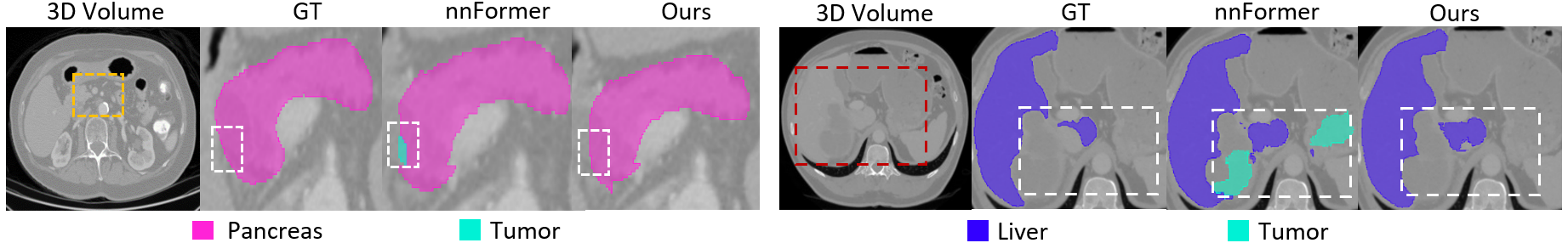}
    \caption{Qualitative results on the  MSD pancreas tumour (left) and  MSD Liver tumour datasets. Our vMixer provides accurate segmentation of boundary regions.}
\label{fig:liver}
\end{figure*}

\begin{table}[t]
\centering
\caption{ {Comparison over in MSD Liver Tumour dataset. The best results are in bold.}} 
\scalebox{0.7}{
\label{tab:msd}
        \begin{tabular}{|c|ccc|ccc|ccc|}
        \hline
        \multicolumn{1}{|l|}{\multirow{2}{*}{Method}} & \multicolumn{3}{c|}{Liver}                                                                & \multicolumn{3}{c|}{Tumor}                                                               & \multicolumn{3}{c|}{Average}                                                              \\ \cline{2-10} 
                                & \multicolumn{1}{c|}{DSC}             & \multicolumn{1}{c|}{HD95}             & NSD           & \multicolumn{1}{c|}{DSC}             & \multicolumn{1}{c|}{HD95}            & NSD           & \multicolumn{1}{c|}{DSC}            & \multicolumn{1}{c|}{HD95}             & NSD            \\ \hline
        3D UNet \cite{3dunet}                & \multicolumn{1}{c|}{94.37}          & \multicolumn{1}{c|}{-}          & -          & \multicolumn{1}{c|}{53.94}         & \multicolumn{1}{c|}{-}         & -         & \multicolumn{1}{c|}{74.15}          & \multicolumn{1}{c|}{-}         & -        \\ \hline  
        UNETR \cite{unetr}                & \multicolumn{1}{c|}{86.69}          & \multicolumn{1}{c|}{20.79}          & 83.06         & \multicolumn{1}{c|}{50.94}         & \multicolumn{1}{c|}{44.68}         &  50.14         & \multicolumn{1}{c|}{68.85}          & \multicolumn{1}{c|}{32.73}         &   66.64      \\ \hline  
        Swin UNETR \cite{swinunetr}                & \multicolumn{1}{c|}{87.31}          & \multicolumn{1}{c|}{19.42}          & 85.07         & \multicolumn{1}{c|}{52.29}         & \multicolumn{1}{c|}{42.96}         &  51.47        & \multicolumn{1}{c|}{69.8}          & \multicolumn{1}{c|}{31.19}         & 68.27         \\ \hline                        
        nnFormer \cite{nnformer}                & \multicolumn{1}{c|}{94.87}          & \multicolumn{1}{c|}{12.77}          & 93.00          & \multicolumn{1}{c|}{55.78}          & \multicolumn{1}{c|}{30.77}         & 60.42          & \multicolumn{1}{c|}{76.75}          & \multicolumn{1}{c|}{21.77}          & 76.72           \\ \hline
        Ours                    & \multicolumn{1}{c|}{\textbf{94.89}} & \multicolumn{1}{c|}{\textbf{10.26}} & \textbf{93.41} & \multicolumn{1}{c|}{\textbf{58.11}} & \multicolumn{1}{c|}{\textbf{28.74}} & \textbf{63.48} & \multicolumn{1}{c|}{\textbf{78.45}} & \multicolumn{1}{c|}{\textbf{19.48}} & \textbf{78.44} \\ \hline
        \end{tabular}}
\end{table}

\begin{table}
\centering
\caption{ {Comparison over  MSD Pancreas Tumour dataset. The best results are in blue}} 
\scalebox{0.7}{
\label{tab:msd_pan}
        \begin{tabular}{|c|ccc|ccc|ccc|}
        \hline
        \multirow{2}{*}{Method} & \multicolumn{3}{c|}{Pancreas}                                                          & \multicolumn{3}{c|}{Tumor}                                                             & \multicolumn{3}{c|}{Average}                                                            \\ \cline{2-10} 
                                & \multicolumn{1}{c|}{DSC}            & \multicolumn{1}{c|}{HD95}           & NSD           & \multicolumn{1}{c|}{DSC}            & \multicolumn{1}{c|}{HD95}           & NSD           & \multicolumn{1}{c|}{DSC}            & \multicolumn{1}{c|}{HD95}            & NSD           \\ \hline
        
        3D-UNet   \cite{3dunet}             & \multicolumn{1}{c|}{69.20}          & \multicolumn{1}{c|}{}          & -          & \multicolumn{1}{c|}{35.64}         & \multicolumn{1}{c|}{-}         & -          & \multicolumn{1}{c|}{52.42}          & \multicolumn{1}{c|}{-}         &     -      \\ \hline
        UNETR \cite{unetr}              & \multicolumn{1}{c|}{71.91}          & \multicolumn{1}{c|}{13.97}          & 86.47      & \multicolumn{1}{c|}{35.93}         & \multicolumn{1}{c|}{26.41}         & 52.92          & \multicolumn{1}{c|}{53.92}          & \multicolumn{1}{c|}{20.19}         &  69.68        \\ \hline
        Swin UNETR \cite{swinunetr}            & \multicolumn{1}{c|}{72.42}          & \multicolumn{1}{c|}{12.65}          & 87.38          & \multicolumn{1}{c|}{36.98}         & \multicolumn{1}{c|}{23.84}         & 53.02         & \multicolumn{1}{c|}{54.70}          & \multicolumn{1}{c|}{18.24}         & 70.23          \\ \hline
        nnFormer \cite{nnformer}               & \multicolumn{1}{c|}{78.80}          & \multicolumn{1}{c|}{6.04}          & 94.91          & \multicolumn{1}{c|}{44.36}         & \multicolumn{1}{c|}{12.83}         & 62.33          & \multicolumn{1}{c|}{61.53}          & \multicolumn{1}{c|}{9.42}         & 78.62          \\ \hline
        Ours                    & \multicolumn{1}{c|}{\textbf{79.81}} & \multicolumn{1}{c|}{\textbf{4.52}} & \textbf{95.92} & \multicolumn{1}{c|}{\textbf{48.45}} & \multicolumn{1}{c|}{\textbf{7.16}} & \textbf{66.32} & \multicolumn{1}{c|}{\textbf{64.13}} & \multicolumn{1}{c|}{\textbf{5.85}} & \textbf{81.12} \\ \hline
        \end{tabular}}   
\end{table}

\begin{table*}[]
\centering
\caption{ {Mean, median, and standard deviation of performance scores over MSD Liver and MSD Pancreas datasets. The best results are in bold.}}
\scalebox{1.0}{
\begin{tabular}{|c|cccccc|cccccc|}
\hline
\multirow{3}{*}{} & \multicolumn{6}{c|}{MSD Liver}                                                                                                                             & \multicolumn{6}{c|}{MSD Pancreas}                                                                                                                                 \\ \cline{2-13} 
                  & \multicolumn{2}{c|}{Mean DSC}                           & \multicolumn{2}{c|}{Median DSC}                         & \multicolumn{2}{c|}{Std Deviation DSC} & \multicolumn{2}{c|}{Mean HD95}                             & \multicolumn{2}{c|}{Median HD95}                           & \multicolumn{2}{c|}{Std Deviation HD95} \\ \cline{2-13} 
                  & \multicolumn{1}{c|}{Liver} & \multicolumn{1}{c|}{Tumor} & \multicolumn{1}{c|}{Liver} & \multicolumn{1}{c|}{Tumor} & \multicolumn{1}{c|}{Liver}   & \multicolumn{1}{c|}{Tumor}   & \multicolumn{1}{c|}{Pancreas} & \multicolumn{1}{c|}{Tumor} & \multicolumn{1}{c|}{Pancreas} & \multicolumn{1}{c|}{Tumor} & \multicolumn{1}{c|}{Pancreas}  & Tumor  \\ \hline
nnFormer    \cite{nnformer}      & \multicolumn{1}{c|}{94.13} & \multicolumn{1}{c|}{55.16} & \multicolumn{1}{c|}{94.25} & \multicolumn{1}{c|}{54.90} & \multicolumn{1}{c|}{0.89}    & 1.06    & \multicolumn{1}{c|}{77.91}    & \multicolumn{1}{c|}{43.14} & \multicolumn{1}{c|}{76.92}    & \multicolumn{1}{c|}{44.11} & \multicolumn{1}{c|}{0.93}      & 1.21   \\ \hline
\textbf{vMixer  (Ours)}          & \multicolumn{1}{c|}{\textbf{94.75}} & \multicolumn{1}{c|}{\textbf{58.08}} & \multicolumn{1}{c|}{\textbf{94.79}} & \multicolumn{1}{c|}{\textbf{58.01}} & \multicolumn{1}{c|}{\textbf{0.76}}    & \textbf{0.96}    & \multicolumn{1}{c|}{\textbf{78.51}}    & \multicolumn{1}{c|}{\textbf{48.27}} & \multicolumn{1}{c|}{\textbf{78.86}}    & \multicolumn{1}{c|}{\textbf{47.96}} & \multicolumn{1}{c|}{\textbf{0.81}}      & \textbf{1.08}   \\ \hline
\end{tabular}
\label{stat}}
\end{table*}

\begin{table}[t]
\caption{Experimental results after end-to-end fine-tuning synapse weights on ZebraFish 3D cell membrane dataset  \cite{cellseg}. We report the results in terms of overall accuracy (JI and DSC), and cell count accuracy (JI/DSC greater than 50\% or 70\%) metrics. Our vMixer performs significantly better against existing methods and achieves state-of-the-art performance. The best results are in bold.}
\scalebox{0.8}{
\begin{tabular}{|l|c|c|c|c|c|c|c|c|}
\hline
           & Avg JI        & Avg DSC       & JI\textgreater{}70\% & DSC\textgreater{}70\% & JI\textgreater{}50\% & DSC\textgreater{}50\% \\ \hline
FocalNet \cite{yang2022focal}          & 51.98          & 62.65         & 39.01                 & 51.7                  & 55.97                 & 71.80                  \\ \hline
ConvNeXt \cite{convnext}         & 52.50          & 64.01         & 39.30                & 53.50                 & 56.09                & 73.60                  \\ \hline
nnFormer\cite{nnformer}     & 52.17          & 63.80         & 38.53                & 53.08                &  \textbf{55.73}                & 73.77                    \\ \hline
Ours            & \textbf{54.25} & \textbf{65.69} & \textbf{42.60}          & \textbf{58.11}         & \textbf{60.91}       & \textbf{76.54}         \\ \hline
\end{tabular}}
\label{hms_finetuning}
\end{table}

\begin{figure*}[]
    \centering
\includegraphics[width=1\textwidth]{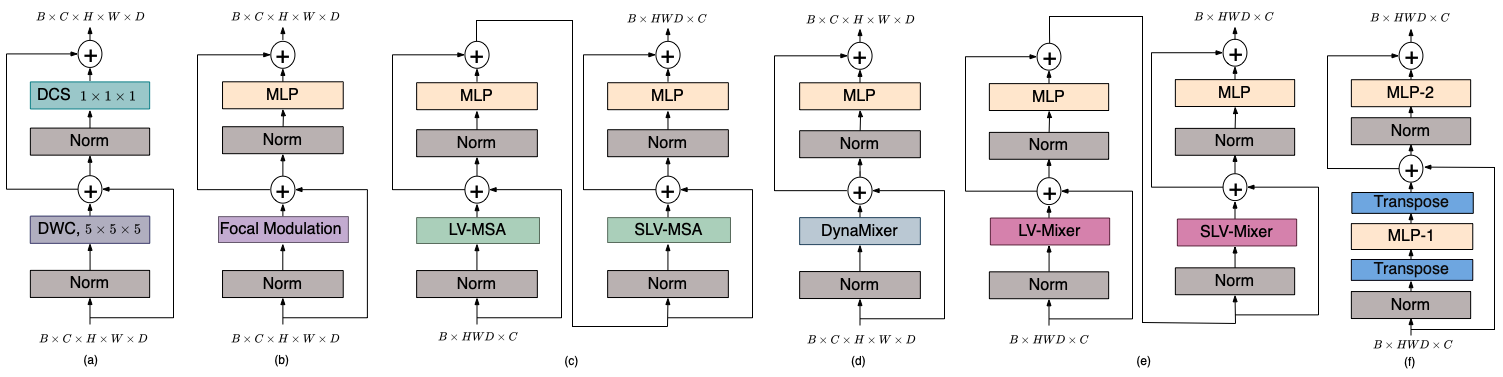}
\caption{\textbf{Adaptation of different network architecture blocks for
3D medical image segmentation}. The depth-wise convolution (DWC) and depth-
wise scaling (DCS) based \textbf{(a) ConvNeXt \cite{convnext}}, \textbf{(b) FocalNet \cite{f_net}} (see Fig.~\ref{fig:q2}-b), and   \textbf{(d) DynaMixer \cite{dynamixer}} (see Fig.~\ref{fig:q2}-a) operate on  input 3D volume of size $B\times C \times H\times W \times D$. 
The  \textbf{(c) Swin Transformer \cite{liu2021swin}},
\textbf{(e) Swin Mixer \cite{liu2021swin}}, and \textbf{(f) GVM} blocks reshape the 3D volume to $B\times HWD \times C$ dimensional features. Here, B: batch, C: channel, H: height, W: width, and D: depth dimensions, respectively.
}
\label{fig:q1}
\end{figure*}

\begin{figure*}[]
    \centering
\includegraphics[width=0.8\textwidth]{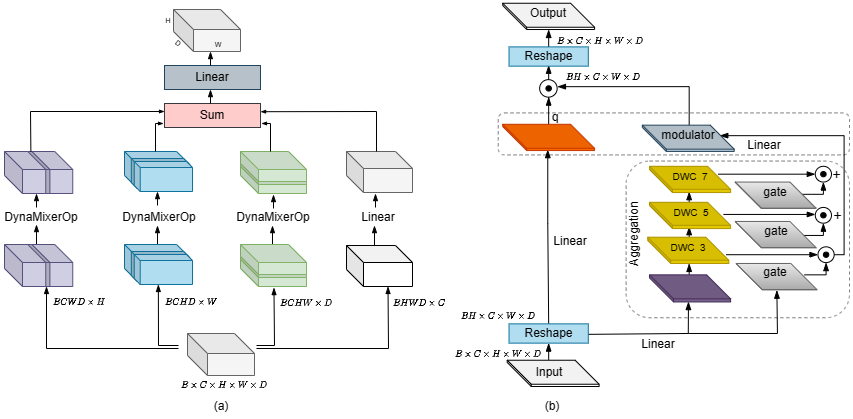}
\caption{\textbf{(a) DynaMixer \cite{dynamixer} Architecture}. The architecture includes height, width, depth, and channel mixing. In height mixing, the tokens are mixed via DynaMixer operations (DynaMixerOp)  along the height dimension using shared weights. Similar operations are performed for width and depth dimensions, respectively. The channel mixing  performs linear transformation of the features. DynaMixer operation (DynaMixerOp) includes mixing elements along the dimension with a series of linear layers followed by a Softmax activation. 
\textbf{(b) Focal Modulation  \cite{f_net} Architecture}. Aggregation consists of hierarchical  gated context aggregation (best viewed in Zoom).}
\label{fig:q2}
\end{figure*}



\begin{table*}[]
\centering
\caption{\textbf{Analysis of various architectures} adapted for 3D medical image segmentation \textbf{in a uniform network where the same architectural block is used in all encoder and decoder stages}  over Synapse Multi-organ dataset. Here, we take representative convolutional,  transformer, and MLP mixer,  architectures and introduce them to a uniform nnFormer-based architecture.  The inferior \textbf{HD95} scores of these \textit{uniform}  architectures compared to the HD95 score obtained by our \textit{hybrid} vMixer (GVM in all stages: 9.90 vs Ours: \textbf{6.78}) demonstrate the benefits of the proposed \textbf{hybrid architecture}  in encoding local and global details for accurate organ segmentation. The structure of these blocks is shown in Fig~\ref{fig:q1}. The best results are in bold}
\label{Analysias_various_architectures}
\begin{tabular}{|ccc|cc|ccc|}
\hline
\multicolumn{3}{|c|}{\textbf{Convolutional Networks}}                   & \multicolumn{2}{c|}{\textbf{Transformers}}  & \multicolumn{3}{c|}{\textbf{Mixers}}                               \\ \hline
\multicolumn{1}{|c|}{3D Conv} & \multicolumn{1}{c|}{ConvNeXt \cite{convnext}} & FocalNet \cite{f_net} & \multicolumn{1}{c|}{Self Attention \cite{dosovitskiy2020image}} & LVSA \cite{nnformer} & \multicolumn{1}{c|}{Swin Mixer \cite{liu2021swin}} & \multicolumn{1}{c|}{DynaMixer \cite{dynamixer}} & GVM \\ \hline
\multicolumn{1}{|c|}{10.78}  & \multicolumn{1}{c|}{32.80}     & 20.66    & \multicolumn{1}{c|}{16.00}           & 10.63 & \multicolumn{1}{c|}{10.10}       & \multicolumn{1}{c|}{11.80}      & \textbf{9.90} \\ \hline
\end{tabular}
\end{table*}

\begin{table*}[h]
\centering
\caption{Ablation Experiments on various \textbf{hybrid architecture} design choices over Synapse Multi-organ dataset. The results show that \textbf{LVSA for capturing local} information at stage 1  and using \textbf{GVM to capture global information} at the remaining stages provides the best result. The best result is in bold. }
\scalebox{0.99}{
\begin{tabular}{|l||c|c||c|c||c|c|c|c|c||c|} \hline 
\textbf{Stage 1} & LVSA & LVSA  & LVSA       & LVSA       & ConvNeXt & FocalNet & DynaMixer & Self-attention & Swin Mixer & LVSA \\\hline
\textbf{Stage 2} & LVSA & LVSA   & Swin Mixer & LVSA       & GVM      & GVM      & GVM       & GVM            & GVM        & GVM  \\\hline
\textbf{Stage 3} & LVSA & GVM    & Swin Mixer & Swin Mixer & GVM      & GVM      & GVM       & GVM            & GVM        & GVM  \\\hline
\textbf{Stage 4 }& GVM  & GVM    & Swin Mixer & Swin Mixer & GVM      & GVM      & GVM       & GVM            & GVM        & GVM  \\ \hline \hline
\textbf{HD95}    & 9.01 &  8.60    & 10.89      & 11.57      & 7.72     &   11.40       & 11.60      &     13.79           &     11.10       & \textbf{6.78} \\ \hline

\end{tabular}}
\label{tab:hyb}
\end{table*}

\begin{figure*}[t]
    \centering
\includegraphics[width=0.8\linewidth]{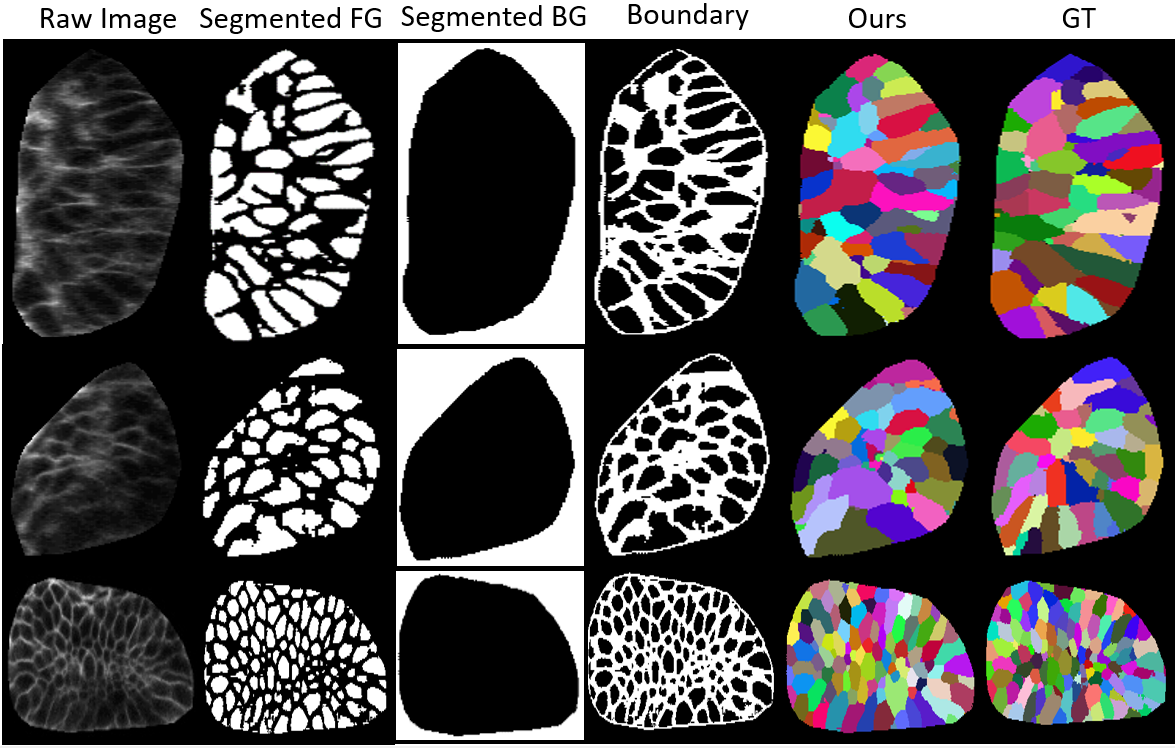}
\caption{Qualitative results of our vMixer  fine-tuned on the Zebrafish 3D cell membrane  dataset  using pre-trained weights from multi-organ Synapse dataset. Rows 1,2 and 3 correspond to views from different planes. The proposed vMixer predicts  foreground (FG),  background (BG), and cell boundary regions (columns 2-4, respectively)  which are post-processed using a watershed algorithm  to accurately segment cell  instances (column~5).}
\label{fig:hms_qualitative}
\end{figure*}

\subsection{State-of-the-art Comparison}
\textbf{\textbf{Synapse Multi-organ dataset:}} In Table \ref{tbl_all_synapse}, we compare our vMixer over {Synapse Multi-organ dataset with the existing state-of-the-art methods. 
MISSFormer \cite{miss}, UNETR \cite{unetr}, swin UNETR \cite{swinunetr} and nnFormer \cite{nnformer} have more than 79\% DSC scores and  HD95 scores are 18.20, 22.97, 17.65 and 10.63, respectively. Although our approach obtains a comparable 86.53\% DSC score, it achieves a better 6.78 HD95 score. This indicates that our method is more capable of capturing the {shape-boundary characteristics} of the organs compared to other SOTA methods.
Furthermore, in table \ref{tb_nnunet}, we conduct a detailed performance analysis between  our method and nnFormer \cite{nnformer} and UNETR \cite{unetr}. Our method shows significant improvement in terms of HD95 scores. 
For example, it can be observed that for the smaller organs with complex boundaries such as the left kidney, right kidney, and gall bladder, our method achieves 5.60, 7.25, and 9.04 HD95 scores. 
In addition, qualitative comparison with other SOTA methods in   \ref{fig:synapse_all}.   It can be clearly seen that our approach provides superior segmentation results compared to other methods and preserves the boundaries for the organs. For example, in figure \ref{fig:synapse_all} (row 1), our method better preserves the shape of the gall bladder organ. 

\textbf{\textbf{MSD Liver Tumour and MSD Pancreas Tumour Datasets:}} We also  compare our method with nnFormer over MSD Liver Tumour and MSD Pancreas Tumour datasets and showed consistent improvement as shown in tables \ref{tab:msd} and \ref{tab:msd_pan}, respectively.  { The qualitative results in figure  \ref{fig:liver} show that our method preserves better {boundary} information.  }

 {In addition, we also perform the statistical analysis to validate the significance of our method, we report the mean, median, and standard deviation of 3-fold experimental results over MSD Liver and MSD Pancreas datasets in Table \ref{stat}. The statistical significance analysis shows that our method clearly outperforms nnFormer \cite{nnformer} in all the quantities analyzed.}

\begin{table}[h]
\centering
\caption{ {Analysis of Transfer Learning experiments on Zebrafish 3D cell membrane Dataset. The best results are in bold.}}
\scalebox{0.8}{
\begin{tabular}{|c|c|c|c|c|}
\hline
\multirow{2}{*}{} & \multicolumn{2}{|c}{Fine-tuning stem and expanding layers} & \multicolumn{2}{|c|}{End-to-end fine-tuning} \\ \cline{2-5} 
    & \multicolumn{1}{c} 
 {    Avg JI}          & Avg DSC         & \multicolumn{1}{c}{   Avg JI  }    & Avg DSC    \\ \hline
FocalNet \cite{f_net}  & \multicolumn{1}{c}{21.96}           & 26.18           & \multicolumn{1}{c}{51.98}     & 62.65     \\ \hline
ConvNeXt   \cite{convnext}       & \multicolumn{1}{c}{24.12}           & 29.42           & \multicolumn{1}{c}{52.5}     & 64.01      \\ \hline
nnFormer   \cite{nnformer}     & \multicolumn{1}{c}{23.89}           & 29.26           & \multicolumn{1}{c}{52.17}     & 63.82      \\ \hline
\textbf{vMixer (Ours)}                 & \multicolumn{1}{c}{\textbf{25.17}}           & \textbf{30.13}           & \multicolumn{1}{c}{\textbf{54.25}}     & \textbf{65.69}      \\ \hline
\end{tabular}
\label{analysis-finetune}
}
\end{table}

\subsection{Transfer Learning: 3D Cell Instance segmentation}
\textbf{ZebraFish 3D cell membrane dataset:} We perform experiments on ZebraFish 3D cell membrane dataset (also named as HMS dataset) to validate the transfer learning abilities of our vMixer on the challenging 3D cell instance segmentation task where accurate boundary prediction is required to delineate different cell instances.  {In this study, we leverage the transfer learning abilities of different architectural choices by taking their pre-trained models from the multi-organ Synapse dataset and fine-tuned the models for the 3D cell instance segmentation task on HMS (ZebraFish 3D cell membrane) dataset having limited training data.
For a fair comparison, we set the same fine-tuning parameters for all architectures, across all our fine-tuning experiments.  We set the batch size as 5,  maximum epochs to 500, and the learning rate as 1e-4 using the Adam optimizer. We set the input size of $64 \times 64 \times 64$ and utilize weighted Dice loss for loss backpropagation. To show the generalizability of our method, we present different transfer learning approaches across various architectural choices over the HMS dataset, as shown in Tab. \ref{analysis-finetune}. Please note that the models trained on the Synapse multi-organ dataset can not be directly utilized on the HMS (ZebraFish 3D cell membrane) dataset due to the following reasons: (i) the input resolution differences (the input resolution for the Synapse multi-organ dataset is $128 \times 128 \times 64$, whereas the input size for the HMS dataset is $64 \times 64 \times 64$). The stem needs to be learned for the target HMS dataset so that the stem can generate appropriate features for the model. (ii) Furthermore, the Synapse Multi-organ dataset has 14 classes and the HMS dataset has 3 classes. Therefore, there is a need to have different expanding layers to handle classes for the HMS dataset. Hence, we take the whole Synapse multi-organ model and make these two minimum modifications to the architecture to adapt to the HMS dataset.}

 {To adapt the multi-organ Synapse weights for the HMS dataset, and show the generalizability of our method, we present two different ways to show the transfer learning abilities of the models over the HMS dataset, as shown in Tab. \ref{analysis-finetune}. 
To compare the generalizability of our model, we first fine-tune only the stem and patch expanding layers by freezing all remaining layers. This minimal learning validates that our frozen model weights have a better generalization ability compared to baseline methods (see Tab. \ref{analysis-finetune} column 1), even without fine-tuning most layers of the multi-organ Synapse pre-trained weights.  Next, we perform end-to-end fine-tuning of the entire network and achieve better JI and DSC scores  (see Tab. \ref{analysis-finetune} column 2). This empirical study reveals that our model has a consistent performance gain,  demonstrating a superior transfer learning ability compared to different architectural choices. Figure \ref{fig:hms_qualitative} shows the qualitative results for an example from ZebraFish 3D cell membrane dataset. We note that our method segments the foreground, background, and boundary of the cells as well as preserves the boundaries for cells. In addition to that, our method provides the 3D cell instance segmentation results. }

\begin{table}
    \centering
\caption{ {Study of vMixer stages suitable for capturing global (GVM)  and local (LVSA) dependencies over Synapse Multi-organ dataset. The best results are in bold.}} 
\scalebox{0.99}{
\label{tab6}
        \begin{tabular}{|c|c|c|c|c|c|}
\hline& Stage 1 & Stage 2 & Stage 3 & Stage 4 &  HD95         \\ \hline
\multirow{2}{*}{\begin{tabular}[c]{@{}c@{}}Uniform \\ Architecture\end{tabular}} & LVSA    & LVSA    & LVSA    & LVSA     & 10.60         \\ \cline{2-6}   & GVM     & GVM     & GVM     & GVM      & 8.60          \\ \hline
\multirow{2}{*}{\begin{tabular}[c]{@{}c@{}}Hybrid \\ Architecture\end{tabular}}  & LVSA    & LVSA    & GVM     & GVM     & 9.90   \\ \cline{2-6}
& LVSA    & GVM     & GVM     & GVM      &  6.78 \\ \hline
\end{tabular}}
\end{table}

                 }

 {\subsection{Discussion of Different Architectural Designs:}}
 {
In the context of accurate segmentation, especially for precise boundary prediction, it is important to thoroughly capture both local and global information. To address this challenge, we meticulously investigate various context aggregation techniques utilized in 2D image classification literature by adapting them to the domain of 3D medical image segmentation.
The CNNs and transformers have already proven to be efficient for volumetric 3D medical imaging as in nnFormer \cite{nnformer} and UNETR \cite{unetr}, the capabilities of MLP-mixer \cite{tolstikhin2021mlp} were not explored for the 3D medical image segmentation tasks.  Moreover, there was little exploration regarding the \textit{hybrid} combinations of multiple context aggregator blocks such convolution, attention, and MLP-mixer for 3D volumetric medical image segmentation. In this work, we strive to explore the inherent characteristics of MLP-mixer based architectures when combined with other context aggregator blocks traditionally used in 2D literature, by adapting them to 3D  in a hybrid design.
Specifically, we explored different context aggregators including CNN-based (3DConv, ConvNeXt \cite{convnext}, and FocalNet \cite{f_net}), transformer-based (self-attention \cite{dosovitskiy2020image}, Swin Transformer \cite{liu2021swin} as LSVA), and mixers-based (Swin Mixer \cite{liu2021swin}, DynaMixer \cite{dynamixer}, MLP-mixer  \cite{tolstikhin2021mlp} as global
volume mixer (GVM)) to learn both local and global context information to learn better features for medical image segmentation. Figure \ref{fig:q1} illustrates the adaptation of different network design blocks.  
}

 {
Our exploration begins with a detailed examination of CNN-based architectures. We observe that traditional 3DConv networks make use of 3D convolutions, which are indeed well-suited for local feature extraction but often prove inadequate in their ability to capture long-range dependencies. We also evaluate the potential of ConvNeXt, an architecture that modernizes conventional convolutional networks by introducing larger kernel sizes and adopting fewer activation functions and normalization layers. We also analyze FocalNets from 2D image recognition literature that employ a focal modulation technique. FocalNets comprise a hierarchical contextualization and a gated aggregation, followed by a modulation strategy. This distinctive approach gathers context information, spanning from short to long-range, and the process of aggregation is influentially guided by the content of the query.}

 {
We also investigate different transformer-based architectures, commencing with the standard self-attention mechanism. While self-attention is capable of encoding long-range dependencies, it may encounter challenges when handling local information. Moreover, it is notably associated with quadratic complexity concerning token length. However, the Swin Transformer enhances locality by adopting a shifted window attention strategy, effectively refining the receptive field while concurrently reducing computational complexity. We evaluate the 3D self-attention and 3D Swin attention in our experiments.}

 {
Finally, we also evaluate MLP mixers-based architectures as context aggregators. MLP-Mixer leverages multi-layer perceptrons to globally mix features, demonstrating its inherent capability to capture complex relationships within the input data. Our proposed Global Volume Mixer (GVM) module is fundamentally derived from the MLP-Mixer architecture, incorporating token-mixing and channel-mixing operations to effectively extract global characteristics from 3D volumetric data. In addition, we also adapt DynaMixer, which is distinguished by its dynamic generation of mixing matrices, predicated on token content analysis, thereby improving computational efficiency and overall robustness. \textit{To the best of our knowledge, this is the first time MLP-mixer based context aggregators are adapted for the problem of 3D medical image segmentation.}}

 {To discern the most suitable architectural design, we thoroughly evaluate the performance of each context aggregator when uniformly applied to all encoder-decoder stages. Significantly, GVM emerges as the top performer, yielding a reduced HD95 score of 9.9, as shown in Tab. \ref{Analysias_various_architectures}.  Further, to optimally capture both local and global information, we explore \textit{hybrid} combinations of these context aggregators. Our experimental results endorse the implementation of LVSA at stage 1, complemented by GVM blocks in subsequent stages, as shown in Tab. \ref{tab:hyb}. Our comprehensive experiments demonstrate the advantages of the proposed hybrid architecture. This configuration, featuring LVSA at the highest resolution and GVM blocks at successive lower-resolution levels, firmly establishes itself as the preferred choice for accurate boundary prediction, as evidenced by the noteworthy improvements in HD95 scores.}

\subsection{Ablation Study}
We perform an ablation study over the Synapse Multi-organ dataset to validate the effectiveness of the proposed vMixer. The proposed vMixer has four encoder stages and three decoder stages. In all our ablation experiments, we use the same architectural design choices for the encoder and decoder stages having the same resolution.  For example,  the first encoder ($E1$) and the last decoder ($D3$) use the same architecture and we refer to this as stage 1. Similarly, the remaining stages (stages 2-4) are also defined based on the respective encoder and decoder stages.  

 {As discussed earlier, we take the different architectures designed for 2D detection tasks and adapt them for 3D medical image segmentation. 
including convolutional network designs (3D Convs, ConvNeXt \cite{convnext}, FocalNet \cite{yang2022focal}), transformers (self-attention \cite{dosovitskiy2020image}, Swin Transformer \cite{liu2021swin} as local volume-based self-attention (LVSA)), and mixers (Swin Mixer as local volume-based MLP-mixer, DynaMixer \cite{dynamixer}, and MLP-mixer \cite{tolstikhin2021mlp} as global volume mixer (GVM)).  
In table \ref{Analysias_various_architectures}, we present an analysis of different architectures adapted for 3D medical image segmentation, where the encoder-decoder has the same architectural block. }
 {We observe that GVM performs better with a reduced HD95 score of 9.90. We also perform an ablation study with various hybrid architecture design choices.  In Table \ref{tab:hyb}, we set different design choices at different stages and observe that our vMixer presents a better HD95 score compared to all other hybrid design choices. Finally, we fix the  GVM in all stages except the first stage of the encoder and the last stage of the decoder, as shown in table \ref{tab:hyb}. We observe that employing   ConvNext or LVSA in the first stage of the encoder and the last stage of the decoder results in superior HD95 scores of 7.72 and \textbf{6.78}, respectively. Based on the aforementioned ablation studies, we fix LVSA at the first stage and GVM at the last three stages of the proposed vMixer.} 

 {Next, we employ different combinations of LVSA and GVM blocks at different stages as shown in Table \ref{tab6}. We can observe that employing GVM  at low-resolution stages helps to better capture global information and optimum performance can be obtained by employing  LVSA (similar to nnFormer) at the first stage and  GVM at all remaining stages (row 3).   The LVSA at stage 1 and GVM blocks at the last 3 stages provide a favorable HD95 score which indicates its better capability to preserve the shapes of different organs. We also present the qualitative results in Figure \ref{vis} and show that our architecture preserves better boundary information. When LVSA is uniformly employed in all stages, the method fails to capture global dependencies and hence loses characteristic details associated with the overall organ shape (as shown in row-2 of Figure \ref{vis}-(ii)).
On the other hand, our study demonstrates that our novel hybrid architecture employing LVSA in the highest resolution and GVM blocks in successive levels helps in a refined feature extraction from the 3D medical volume aiding improved boundary detection and facilitating better segmentation.}

\begin{figure}
 \centering
\caption{ {Comparison between nnFormer \cite{nnformer} and vMixer with respect to the number of epochs over Synapse Multi-organ dataset.}}   
    \includegraphics[width=1.0\linewidth]{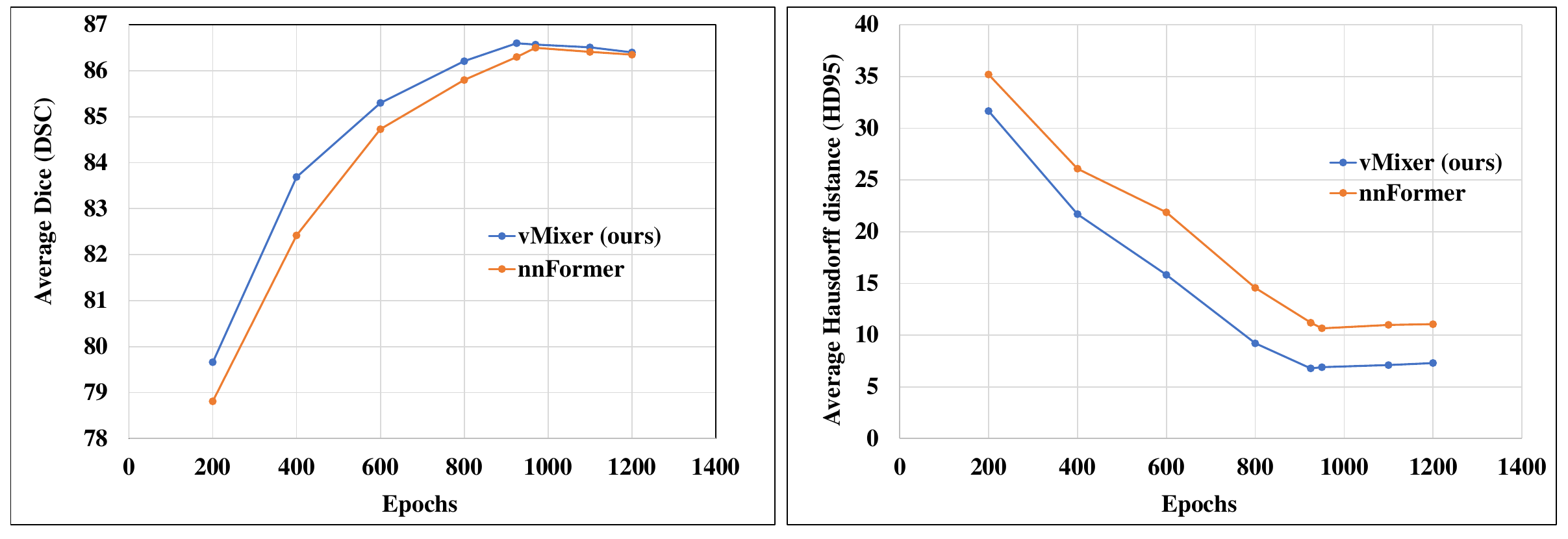}
     \label{convergence_comparison}
\end{figure}

 {\textbf{Training Epochs vs Model Convergence:}}
 {We perform experiments with 1000 maximum epochs to be consistent with our baseline (nnFormer \cite{nnformer}). We have extended the experiments by increasing the number of epochs to 1200  as shown in Figure \ref{convergence_comparison}. It can be seen that our method has a clear dominance in performance scores and achieves faster convergence compared to the nnFormer, especially in terms of HD95 score  (as shown in Fig. \ref{convergence_comparison}-right). We observe that there is enormous performance improvement for both nnFormer and vMixer until 900 training epochs, and later, optimal performance is achieved between the 900th and 1000th epoch. Thereafter, the model performance did not undergo a noticeable change, and hence the total number of epochs was set to 1000. We also observe from Fig. \ref{convergence_comparison} that training for fewer epochs may not be conclusive about the exact performance trend as 3D datasets are complex and demand larger network architectures.  
}

 {\textbf{Computational Cost Analysis:}}
 { We present the computational analysis of our method. We present the comparison of floating-point operations per second (FLOPs) and the inference time. The computational cost comparison for different methods over a multi-organ synapse dataset is presented in Table \ref{flop}. 
It can be seen that our vMixer obtains a prominent dominance over other SOTA methods in terms of HD95 scores with comparable FLOPs and inference time. Therefore, there is a tradeoff between the accuracy and inference time. 
}
\begin{table}[]
\centering
\caption{ {Computational cost comparison over Synapse Multi-organ  dataset.}}
\scalebox{1.1}{
\begin{tabular}{|c|c|c|c|}
\hline
\textbf{Model} & \textbf{Flops} & \textbf{Inference time (ms)} &  \textbf{HD95}\\ \hline 
swin UNETR \cite{swinunetr}                  & 572        &   228.6& 17.65\\ \hline 
nnFormer \cite{nnformer}                   & 212        &   148.0  & 10.6   \\ \hline 
vMixer                       & 249        &   154.3  & 6.78 \\ \hline 
\end{tabular}}
\label{flop}
\end{table}

\section{Conclusion}
We propose a hierarchical encoder-decoder network to explicitly learn the local and global dependencies. We utilize the local volume-based self-attention to learn the local dependencies at high-resolution features and propose a volumetric global mixing mechanism to capture the global feature representations at low-resolution features.  These explicit local-global feature representations benefit to capture the boundaries of the organs. Experimental study reveals that our approach provides favorable segmentation results at boundary regions  compared to existing SOTA methods. Moreover, our experiments show that our vMixer provides promising 3D cell instance segmentation results on   the Zebrafish cell 3D instance segmentation dataset.


\section*{Acknowledgement}
This work is partially supported by the MBZUAI-WIS Joint Program for AI Research (Project grant number- WIS P008).

\ifCLASSOPTIONcaptionsoff
  \newpage
\fi

\bibliographystyle{IEEEtran}
\bibliography{IEEEabrv, egbib}

\end{document}